\documentclass[twocolumn,showpacs,superscriptaddress,preprintnumbers,amsmath,amssymb,prc]{revtex4-2}

\usepackage{graphicx}
\usepackage{dcolumn}
\usepackage{bm}
\usepackage{indentfirst} 
\usepackage{color}
\usepackage{CJK}
\usepackage{booktabs}
\usepackage{float}
\usepackage[colorlinks,linkcolor=blue,urlcolor=blue,citecolor=blue]{hyperref}
\usepackage{tikz,xcolor,hyperref}

\definecolor{lime}{HTML}{A6CE39}
\DeclareRobustCommand{\orcidicon}{
	\begin{tikzpicture}
	\draw[lime, fill=lime] (0,0) 
	circle [radius=0.16] 
	node[white] {{\fontfamily{qag}\selectfont \tiny ID}};
	\draw[white, fill=white] (-0.0625,0.095) 
	circle [radius=0.007];
	\end{tikzpicture}
	\hspace{-2mm}
}
\foreach \x in {A, ..., Z}{%
	\expandafter\xdef\csname orcid\x\endcsname{\noexpand\href{https://orcid.org/\csname orcidauthor\x\endcsname}{\noexpand\orcidicon}}
}
\foreach \x in {A, ..., Z}{%
	\expandafter\xdef\csname orcid\x\endcsname{\noexpand\href{https://orcid.org/\csname orcidauthor\x\endcsname}{\noexpand\orcidicon}}
}

\begin{document}

\begin{CJK}{UTF8}{gbsn}
\title{Branching ratio and information entropy for the p+$^{12}$C reaction in the extended-quantum-molecular-dynamics model}

\author{Lei Shen(沈雷)}
\affiliation{Shanghai Research Center for Theoretical Nuclear Physics， NSFC and Fudan University, Shanghai 200438, China}
\affiliation{Shanghai Institute of Applied Physics, Chinese Academy of Sciences, Shanghai 201800, China}
\affiliation{School of Physical Science and Technology, ShanghaiTech University, Shanghai 201203, China}
\affiliation{University of the Chinese Academy of Sciences, Beijing 100080, China}

\author{Bo-Song Huang(黄勃松)}
\affiliation{Shanghai Institute of Applied Physics, Chinese Academy of Sciences, Shanghai 201800, China}
\affiliation{Shanghai Research Center for Theoretical Nuclear Physics， NSFC and Fudan University, Shanghai 200438, China}

\author{Yu-Gang Ma(马余刚)\orcidB{}}
\email{mayugang@fudan.edu.cn}
\affiliation{Shanghai Research Center for Theoretical Nuclear Physics， NSFC and Fudan University, Shanghai 200438, China}
\affiliation{Key Laboratory of Nuclear Physics and Ion-Beam Application (MOE), Institute of Modern Physics, Fudan University, Shanghai 200433, China}

\date{\today}
\begin{abstract}
The reactions of p+$^{12}$C with triangular 3$\alpha$ structure and spherical structure with incident energies ranging from 5 to 200 MeV/nucleon are simulated by the in the extended-quantum-molecular-dynamics  (EQMD) model together with the GEMINI decay process. This paper presents the incident energy dependence of the branching ratios of fragment production and multiplicity, as well as the associated event information entropy. The results indicate that the triangular 3$\alpha$ $^{12}$C has an extra branching ratio for the quasi-elastic reaction compared to the spherical $^{12}$C. This peculiarity of the branching ratios of the triangular 3$\alpha$ $^{12}$C appears as a small dent in the curves of the event information entropy (both the fragment information entropy and the multiplicity information entropy). Therefore, we propose that the event information entropy could be a probe for detecting the $\alpha$-cluster structure.
\end{abstract}

\maketitle

\section{Introduction}
The multi-particle production through branching process is a kind of chaotic process \cite{ZC1,ZC2}, where the same initial state can lead to different final states upon observation. The entropy of a chaotic system is an important property to evaluate its chaoticity. The multi-particle production through branching process is considered as a dynamical system in which the entropy generally increases. The information entropy has been used to study the chaoticity of such system \cite{PB1,YGM1,CWM1}. The concept of information entropy was originally defined by Shannon \cite{Shannon} and can be expressed as follows
\begin{equation}
H = - \sum_{i}p_i ln(p_i), 
\label{eq_ie}
\end{equation}
where $p_i$ is a normalized probability of a certain physical observable, and $\sum_i p_i =1$. 
In previous studies of nuclear physics, different definitions of information entropy have been studied, such as event information entropy \cite{YGM1,GLM1,JP1} and configurational information entropy \cite{WK1}. The idea of event information entropy in heavy-ion collisions was first proposed by Ma to study the nuclear liquid-gas phase transition \cite{YGM1}. Recently, the information entropy for the Au + Au collision in relativistic heavy ion collisions has also been studied, showing that it is indicative of QCD phase changes \cite{Deng3}, and the diversity of nuclear density can also be studied by the information entropy \cite{MaWH}.

To describe the evolution of heavy-ion collisions, transport models have been widely used \cite{Wolter,Ono2006EPJA,Ono2019PPNP,Xu2019PPNP,Deng1}. There are basically two families of transport models, namely the Boltzmann-Uehling-Uhlenbeck (BUU) models and the Quantum Molecular Dynamics (QMD) models. The BUU models, which is based on the test particle method, cannot describe the event-by-event fluctuation clearly due to the challenge of identifying different fragments. In contrast, the QMD models can more naturally describe the event-by-event fluctuations by generating different events that evolve independently according to branching processes.

The QMD models are particularly effective for the description of fragmentation process in heavy-ion collisions \cite{Aichelin}. Numerous applications for different research topics have been presented in the fields in past decades, eg. see Refs.~\cite{Deng2,Zhou,Qiao,Ding,Buy,Cao,WangSS,Nara,WangTT,WangFY,XiaoK,WangRS,LiuC,LiL,LiPC1,LiPC2,Ding2}. Different from the original QMD model, the extended quantum molecular dynamics (EQMD) model was proposed by Maruyama {\it et al.} with several improvements \cite{Maruyama}, and it can well describe the properties of nuclear ground state, $\alpha$-cluster structure and even halo structure of light nuclei \cite{W.B.He,W.B.He2,Huang2021,YeYL,MaYG,MaYG2}.

In the present work, we use the EQMD model to study the effects of different initial states on branching ratios of fragment production and multiplicity as well as their information entropies. We focus on the p + $^{12}$C reaction with incident energies ranging from 5 to 200 MeV/nucleon, and the reactions with the target nucleus $^{12}$C with 3$\alpha$-cluster structure or spherical structure are simulated separately.

The paper is organized as follows. Sec. II gives a brief introduction to the EQMD model. 
Sec. III gives results and discussion for the p + $^{12}$C reaction for incident energies from 5 to 200 MeV using the EQMD model plus the GEMINI decay code. In this section, both the fragment branching ratio and the multiplicity 
branching ratio as well as their information entropies are presented and discussed.
Comparisons between the results with EQMD only and EQMD + GEMINI are also discussed. Finally, a summary of the work is given in Sec. IV.

\section{EQMD Model}

In the EQMD model, a nucleon is described by a complex Gaussian wave packet as following,
\begin{multline}
\phi_{i}(\vec{r}_{i}) = \bigg(\frac{v_i+v^{*}_{i}}{2\pi}\bigg)^{3/4}exp\bigg[-\frac{v_{i}}{2}(\vec{r}_{i}-\vec{R}_{i})^{2} +\frac{i}{\hbar}\vec{P}_{i}\cdot \vec{r}_{i}\bigg],
\label{Eqa_rhor}
\end{multline}
where $\vec{R}_{i}$ and $\vec{P}_i$ are the centers of the wave packet in coordinate and momentum spaces, and $v_i=\frac{1}{\lambda_i}+i\delta_i$ is the complex wave packet width corresponding to the $i$-th nucleon. 

The interaction potential in the EQMD model consists of four terms, namely Skyrme potential ($\mathcal{H}_{Skyrme}$), Coulomb potential ($\mathcal{H}_{Coulomb}$), symmetry energy potential ($\mathcal{H}_{Symmetry}$) and Pauli potential ($\mathcal{H}_{Pauli}$), and it is written as 
\begin{equation}
\mathcal{H}_{int} = \mathcal{H}_{Skyrme} + \mathcal{H}_{Coulomb} + \mathcal{H}_{Symmetry} + \mathcal{H}_{Pauli}.
\end{equation}
The followings are the different interaction terms, 
\begin{equation}
\mathcal{H}_{Skyrme} = \frac{\alpha}{2\rho_0}\int\rho^2(\vec{r})d^{3}r+\frac{\beta}{(\gamma+1)\rho_0^\gamma}\int\rho^{\gamma+1}(\vec{r})d^{3}r,
\end{equation}
\begin{equation}
\mathcal{H}_{symmetry} = \int\frac{C_s}{2}\frac{(\rho_p-\rho_n)^{2}}{\rho_0}d^3r,
\end{equation}
\begin{equation}
\mathcal{H}_{Pauli} = \frac{C_P}{2}\sum_i(f_i-f_0)^{\mu}\theta(f_i-f_0),
\label{eq:pauli}
\end{equation}
where $C_P$ and $\mu$ are the strength and power of the Pauli potential, $\theta$ is the unit step function, $f_i$ is the overlap of a nucleon with the same spin and isospin nucleons including itself, i.e. $f_i \equiv \sum_j \delta\left(S_i, S_j\right) \delta\left(T_i, T_j\right)\left|\left\langle\phi_i \mid \phi_j\right\rangle\right|^2$, and $f_0$ is the threshold parameter, which takes a value close to 1.
When the $f_0=1$, the step function in Eq.~\ref{eq:pauli} can be ignored, since the $f_i-f_0$ is always greater than 0.
The Pauli term can be understood as a repulsive force forbidding the nearby identical particle too close in the phase space. It makes the EQMD capability to describe $\alpha$ clustering structure in a nucleus.

For the treatment of two-body collisions, the prescription of the standard QMD is employed in the EQMD \cite{Maruyama}. If a pair of two nucleons fulfill these conditions, i.e., (1) their relative distance takes its minimum value within the time-step, and (2) the minimum distance being smaller than a certain value $d_{coll}$, then a stochastic two-body collision is set to occur (3) with the probability $P_{coll}$ decided as
\begin{equation}
P_{coll}=\frac{\sigma_{NN}}{\pi d_{coll}^2},
\label{coll3}
\end{equation}
where $d_{coll}$ is taken 2.0 fm and the nucleon-nucleon cross section is written as
$\sigma_{NN} = \frac{100}{1+\epsilon/200}$
with $\epsilon\equiv\frac{{p_{rel}}^2}{2m}$.

The fragments produced by the EQMD model at the end of the finite simulation time (300 fm/c in the present work) are moving independently, and they may be unstable nuclei or in excited states, which are treated by a statistical decay process called the GEMINI model \cite{Charity2008INDC}. Using information about a given primary fragment, including its proton number, mass number, excitation energy, and spin, the GEMINI model de-excites the fragment through a series of sequential binary decays, i.e., light-particle evaporation, fission, and gamma-ray emission, until the excitation energy of the fragments reaches zero.

\section{Results and discussion}

\begin{figure}[htp]
  \includegraphics[width=\columnwidth]{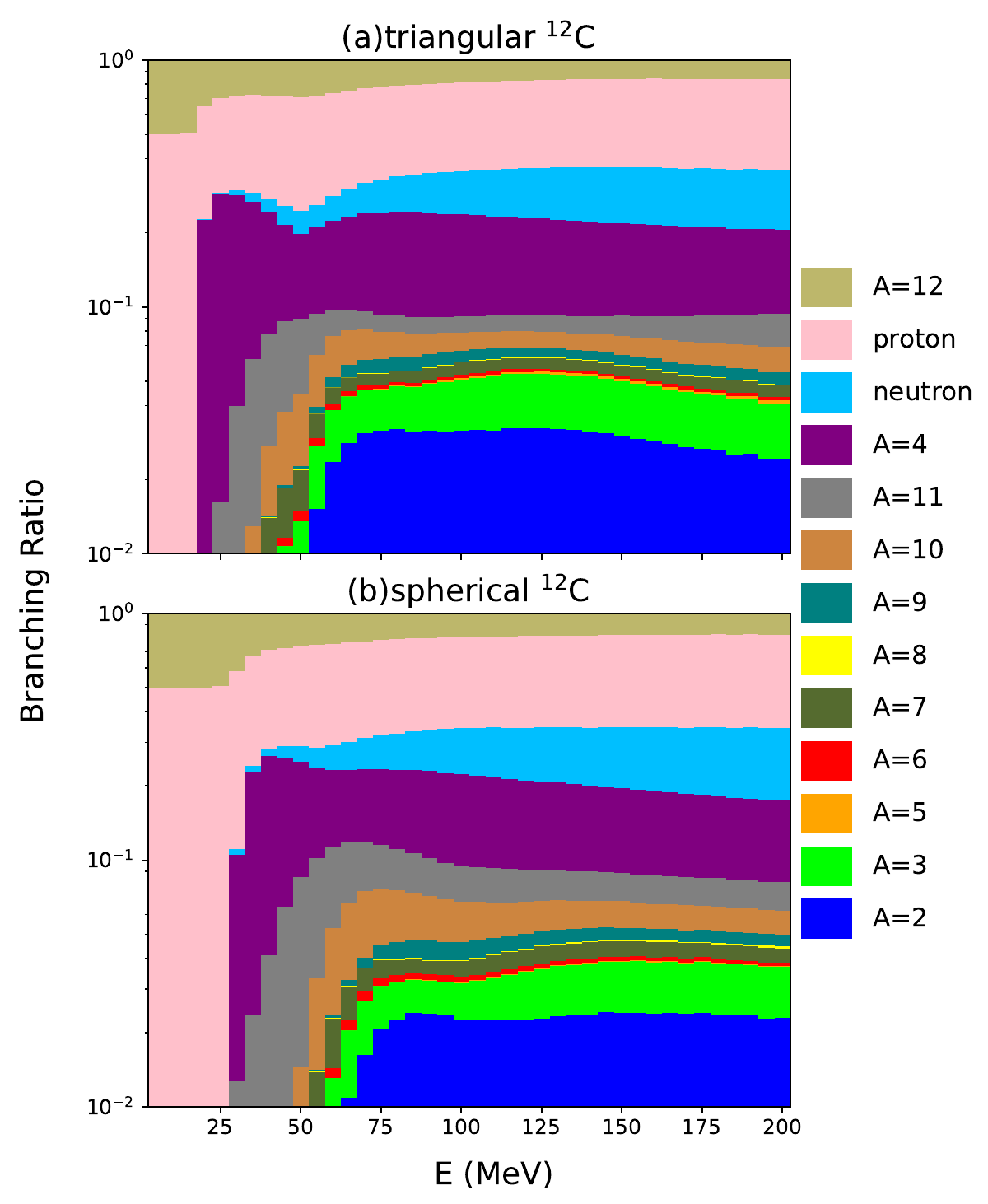}
  \caption{The distribution of the fragment branching ratio $p_i$ of proton, neutron and fragments with mass number from 2 to 13 as a function of the incident energy for the p + $^{12}$C reaction. Panel (a) shows the result of target $^{12}$C with triangular 3$\alpha$-cluster structure, while panel (b) shows the result of target $^{12}$C with spherical structure. Both panels are the results by the EQMD model with GEMINI decay process.}
  \label{fig_Asum}
\end{figure}

Two kinds of $^{12}$C structure are generated by EQMD model \cite{W.B.He}. One is triangular 3$\alpha$-cluster structure with binding energy $E_{\text{bind}} = 87.1$ MeV, and the other is the Wood-Saxon distribution with binding energy $E_{\text{bind}} = 104.5$ MeV, which is referred to as the spherical distribution in this work. Proton induced reactions are commonly used as a probe of nuclear structure, and thus we used the EQMD model to simulate the reaction of a proton with beam energy ranging from $5$ to $200$ MeV colliding with the $^{12}$C target with different structures and impact parameters ranging from $0$ to $4$ fm. The event number for the EQMD simulation for each structure of $^{12}$C is $10^6$, and each emitted fragment undergoes ten simulations of decay by the GEMINI model.

\subsection{Frangment branching ratio}\label{sec_br}

The fragment branching ratio $p_i$ is generally defined as,
\begin{equation}
  p_i = \frac{N_i}{N_0}
  \label{eq_br}
\end{equation}
where $N_i$ is the number of fragment $i$, and $N_0$ is the number of all fragments, and thus $p_i$ refers to the normalized production probability of fragment $i$.

\begin{figure*}[htp]
  \includegraphics[width=\textwidth]{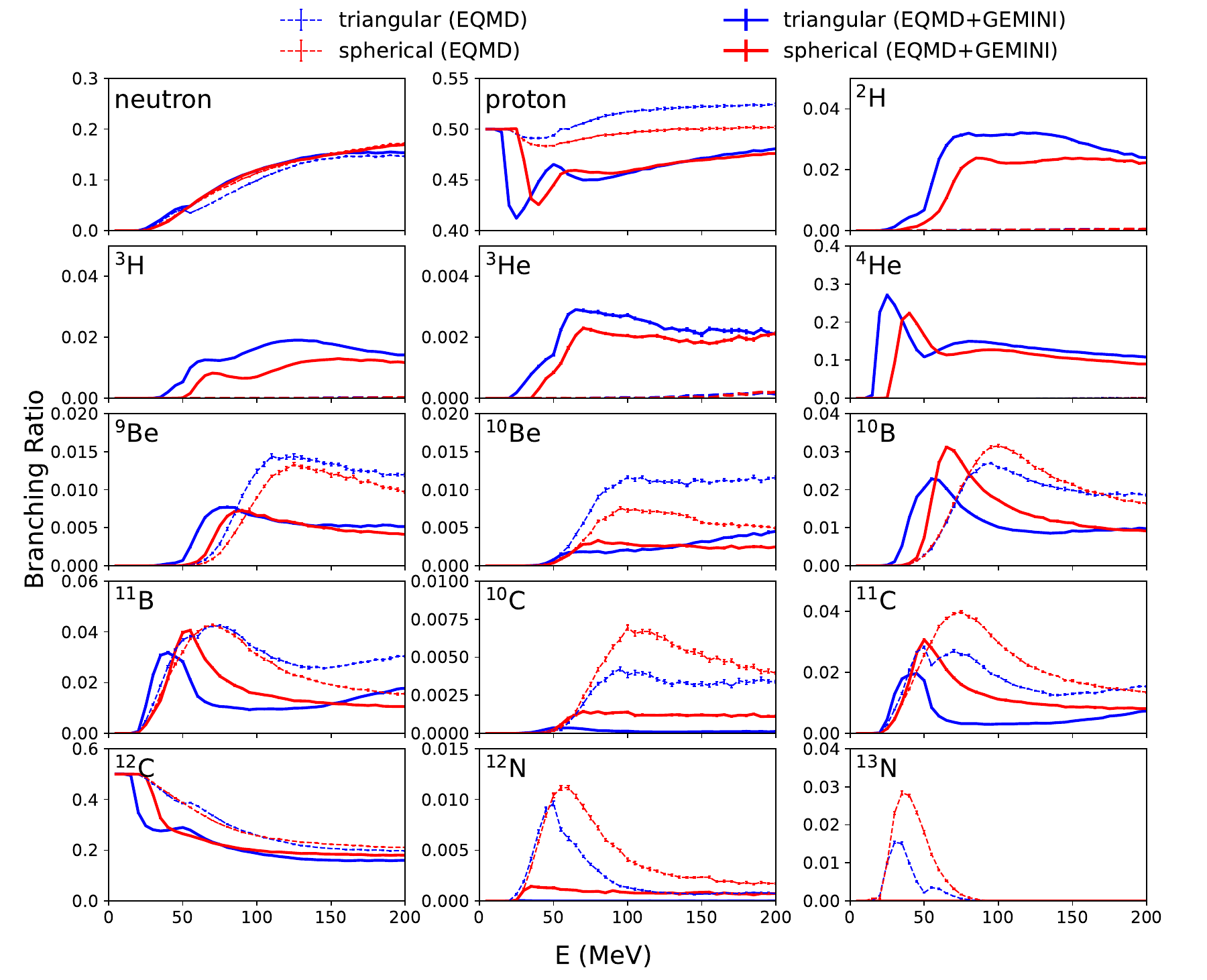}
  \caption{The incident energy dependence of the fragment branching ratio $p_i$ of different fragments in the p+$^{12}$C reaction. The blue lines are the results of the target $^{12}$C with triangular 3$\alpha$ cluster structure, while the red lines are the results of the target $^{12}$C with spherical structure. The solid lines are the results of the EQMD model with GEMINI decay process, while the dashed lines are the results of the EQMD model only.}
  \label{fig_branch}
\end{figure*}

First, to show the global character of the fragmentation, Fig.~\ref{fig_Asum} shows the distribution of the fragment branching ratios as a function of the incident energy $E$. It can be seen that $\alpha$ particle production ($A=4$) is a dominant fragmentation channel in addition to the quasi-elastic reaction products, and the single-nucleon removal reaction ($A=11$) also shows a significant branching ratio. For a more detailed comparison, the main fragments are shown respectively in Fig.~\ref{fig_branch}. The results from the EQMD model only (dashed lines), which can be understood as direct dynamics results, are also shown in Fig.~\ref{fig_branch}. This is because we believe that the direct dynamics results are more sensitive to the nuclear structure, due to the loss of part of the nuclear structure information during the GEMINI decay process. In particular, the nuclear structure only affects the excitation energy in the GEMINI model input quantities, which means that the information of a spatial distribution is simplified to a single value. Furthermore, the direct dynamics results can help us to better understand the reaction mechanisms.

As shown in Fig.~\ref{fig_branch}, the branching ratios of neutrons and protons are associated with a mixture of multiple reaction channels. The neutron branching ratio shows an increasing trend, while the proton branching ratio appears more complex. The branches for almost all light fragments, such as $^{2}$H, $^{3}$H, $^{3}$He, and $^{4}$He, originate from GEMINI decay processes. In contrast, the branching ratios for heavier fragments, such as $^{11}$B, $^{11}$C, $^{12}$N, and $^{13}$N, are generally reduced after the decay processes. In addition, each heavy fragment corresponds to a specific reaction channel, which will be discussed separately below.

In Fig.~\ref{fig_branch}, the panels for $^{2}$H, $^{3}$H, $^{3}$He, and $^{4}$He show that the energy threshold for spherical $^{12}$C are shifted to the higher energy side by about 15 MeV compared to that for triangular $^{12}$C, due to the binding energy difference between two structures of $^{12}$C. The curve shapes of the two structures are very similar because the structural information is not fully respected in the decay process.

In Fig.~\ref{fig_branch}, the panel of $^{13}$N refers to the fusion reaction. All the $^{13}$N fragments have been destroyed after the decay processes. According to the EQMD model results (dashed lines), the fusion branching ratio for spherical $^{12}$C is larger than that for triangular $^{12}$C. This suggests that a nucleus with a more uniform distribution may have a larger absorption cross section because it can more effectively absorb the kinetic energy of the incoming nucleon. In addition, since the inner $\alpha$-cluster is relatively stable, it is less likely to capture an additional proton to form a $^5$Li-like structure in the nucleus.

In Fig.~\ref{fig_branch}, the panels of $^{11}$B and $^{11}$C refer to the single-nucleon knockout reaction. For both structures of $^{12}$C it can be seen that the branching ratios of $^{11}$B and $^{11}$C with the decay process (solid lines) are generally lower than those without the decay processes (dashed lines), which means that a considerable part of the products of the direct dynamics results are in the excited state and later decay into other fragments. Only in the lower energy part (about $E<50$ MeV) the branching ratios of $^{11}$B and $^{11}$C with the decay process (solid lines) are higher than those without the decay processes (dashed lines). This suggests that single nucleon evaporation from an excited $^{12}$C contributes a lot in the lower energy part, so that the difference between two structures of $^{12}$C in this energy range (about $E<50$ MeV) mainly depends on the difference in binding energy due to different structures of $^{12}$C. On the other hand, for the higher energy part (about $150<E<200$ MeV), the branching ratios of $^{11}$B and $^{11}$C of $^{12}$C with different structures show opposite trends. In particular, the branching ratios of $^{11}$B and $^{11}$C of the triangular $\alpha$ cluster $^{12}$C (blue solid lines) both show an increasing trend. In contrast, the branching ratios of $^{11}$B and $^{11}$C of spherical $^{12}$C (red solid lines) both show a decreasing trend. The same trend difference can also be observed in the direct dynamics results (dashed lines). This suggests that in the higher energy range (about $150<E<200$ MeV) the direct kinetic processes (EQMD simulations) dominate the single-nucleon removal reactions. Such a feature can also be observed in the branching ratios of $^{10}$B and $^{10}$Be. Another interpretation of this feature is that, compared to the results for spherical $^{12}$C, the result for triangular $^{12}$C shows a loss of yield mainly in the middle energy range (about $50<E<100$ MeV), and this loss decreases as the incident energy increases further. A possible explanation for this interpretation is that this loss transitions into an excess in the branching ratio of the quasi-elastic reaction for the triangular $^{12}$C compared to the spherical $^{12}$C, as shown in the panel of $^{12}$C in Fig.~\ref{fig_branch}.

In Fig.~\ref{fig_branch}, the panel of $^{12}$N refers to the nucleon transfer reaction. Although the direct dynamics results of both structures of $^{12}$C (dashed lines) have $^{12}$N products, after the decay process only the result of spherical $^{12}$C (red solid line) has a small branching ratio of $^{12}$N. The lower energy part (about $20<E<80$ MeV) may be explained by the evaporation of the exited $^{13}$N, while the higher energy part (about $E>80$ MeV) must be the stable $^{12}$N directly produced by the direct dynamics evolution, since there are no excited $^{13}$N produced for $E>80$ MeV. This result indicates that neutron-proton exchange is much easier in the uniform spherical distribution of $^{12}$C, which can be attributed to the looser neutron and proton distribution compared to the more tightly bound inner $\alpha$-clustering structure of $^{12}$C. The panel of $^{10}$C also shows a similar feature to the panel of $^{12}$N, that only the result of spherical $^{12}$C (red solid line) has a small branching ratio of $^{10}$C after the decay process. This suggests that the nucleon transfer reaction dominates the production of $^{10}$C rather than the two-neutron knockout reaction.

In Fig.~\ref{fig_branch}, the panel of $^{12}$C refers to the quasi-elastic reaction, which does not produce any new nucleus. For both the direct dynamics results and the decay results of the triangular $^{12}$C (blue dashed line and blue solid line), a bump can be observed on the curve of branching ratio, in contrast to the smooth curve of the spherical $^{12}$C. This result suggests that the $\alpha$-clustering structure of $^{12}$C is more prone to quasi-elastic reaction in this energy range (about $40<E<80$ MeV). The position of the bump in the direct dynamics result for triangular $^{12}$C (blue dashed line) also corresponds to the inflection points at $E\approx 50$ MeV in the blue dashed lines in the panels for $^{11}$B, $^{11}$C, and $^{13}$N. Such inflection points can be understood as the peak of the extra quasi-elastic branching ratio distribution of the triangular $^{12}$C compared to the spherical $^{12}$C case. Since the quasi-elastic reaction is dominated by the direct dynamics process (EQMD model), this peak is sharply distributed in the direct dynamics results (blue dashed lines), and the decay process broadens it, resulting in a smoother distribution (blue solid lines). Such an explanation is consistent with our discussion on the panels of $^{11}$B and $^{11}$C above.

\begin{figure}
  \includegraphics[width=\columnwidth]{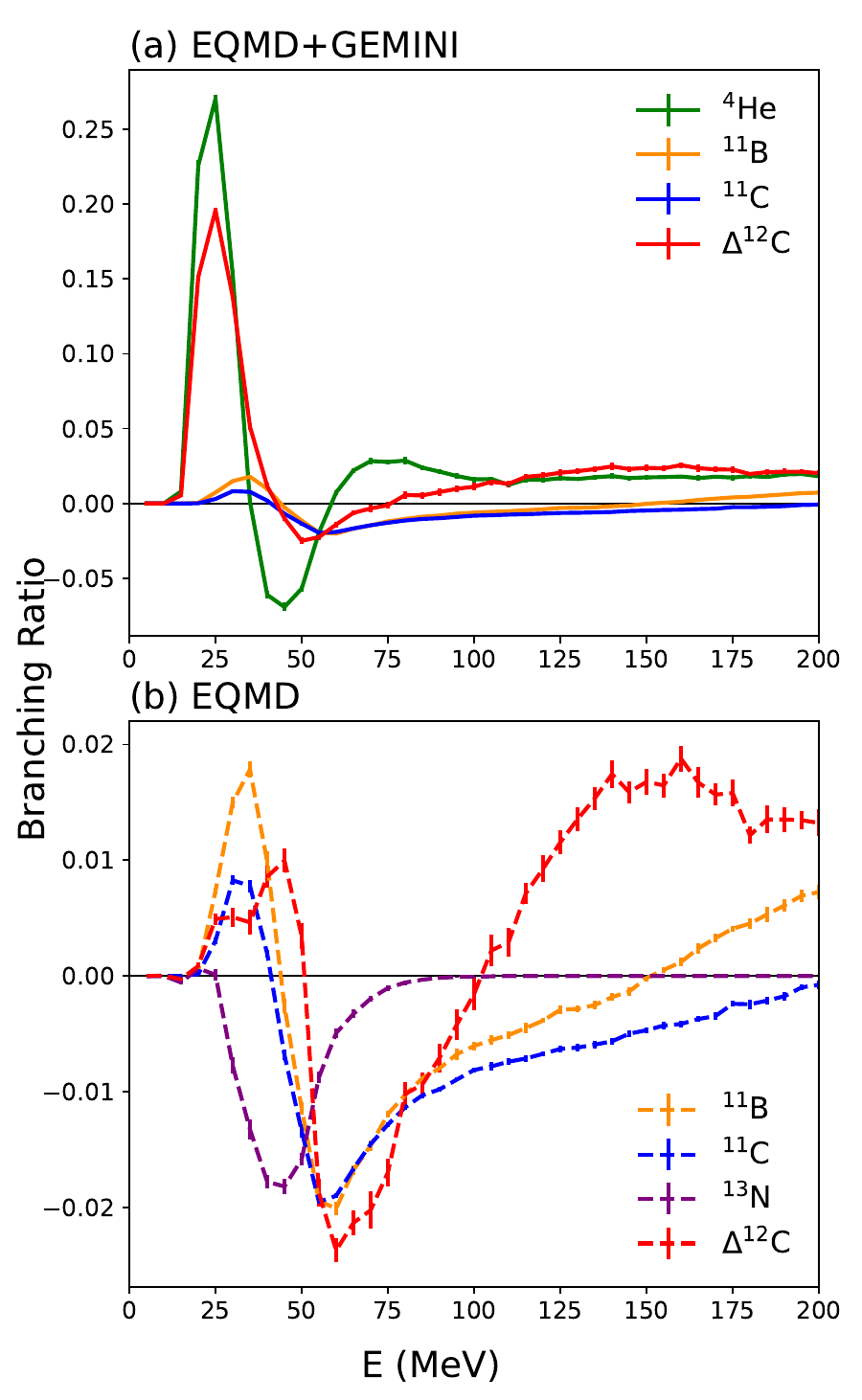}
  \caption{The difference in the branching ratios of some main fragments between the two structures, as a function of the incident energy. Specifically, $p(\text{diff})=p(\text{tri})-p(\text{sph})$. For example, the green solid line here is the result of the blue solid line minus the red solid line in the panel for $^{4}$He in Fig.~\ref{fig_branch}. Panel (a) is the result after the GEMINI decay process and shows the results of $^{4}$He (green solid line), $^{11}$B (yellow solid line) and $^{11}$C (blue solid line). Panel (b) is the result by the EQMD model only and shows the results of $^{11}$B (yellow dashed line), $^{11}$C (blue dashed line) and $^{13}$N (purple dashed line). Both panels show the result of the reduced branching ratios of $^{12}$C compared to the initial state (red lines), which is $p(\Delta^{12}\text{C})=0.5-p(^{12}\text{C})$.}
  \label{fig_brdif}
\end{figure}

Fig.~\ref{fig_brdif} shows the differences in fragment branching ratios caused by different initial nuclear structures. The red lines in both panels are the reduced branching ratios of $^{12}$C compared to the initial state, which is related to the total inelastic collision cross section. In panel (a) for the decay results, both curves for $^{4}$He and $\Delta^{12}\text{C}$ show a high sharp peak at $E\approx 25$ MeV and a valley at $E\approx 50$ MeV resulting from the decay threshold shift due to the different binding energies of the two $^{12}$C structures. The curves for $^{11}$B and $^{11}$C are almost the same, and their values are relatively small compared to the curves for $^{4}$He and $\Delta^{12}\text{C}$. Thus, panel (a) indicates that the production of $^{4}$He, which is mainly due to the 3-$\alpha$ decay of excited $^{12}$C, dominates the structural difference in the branching ratios. On the other hand, in panel (b) for the direct dynamics results, the curves for $^{11}$B, $^{11}$C and $\Delta^{12}\text{C}$ all show a deep valley at $E\approx 55$ MeV, which corresponds to the peak of the extra quasi-elastic branching ratio distribution of the triangular $^{12}$C discussed above. Although the blue dashed line in the panel for $^{13}$N in Fig.~\ref{fig_branch} shows a valley at $E\approx 50$ MeV, it is not prominent here. The result for $^{13}$N (purple dashed line) is mainly due to the difference in proton absorption cross sections between different structures, rather than the difference in quasi-elastic branching ratios caused by different structures. Thus panel (b) shows that the single-nucleon removal reaction dominates the structure difference in the branching ratios in the direct dynamics stage.

\begin{figure}
  \includegraphics[width=\columnwidth]{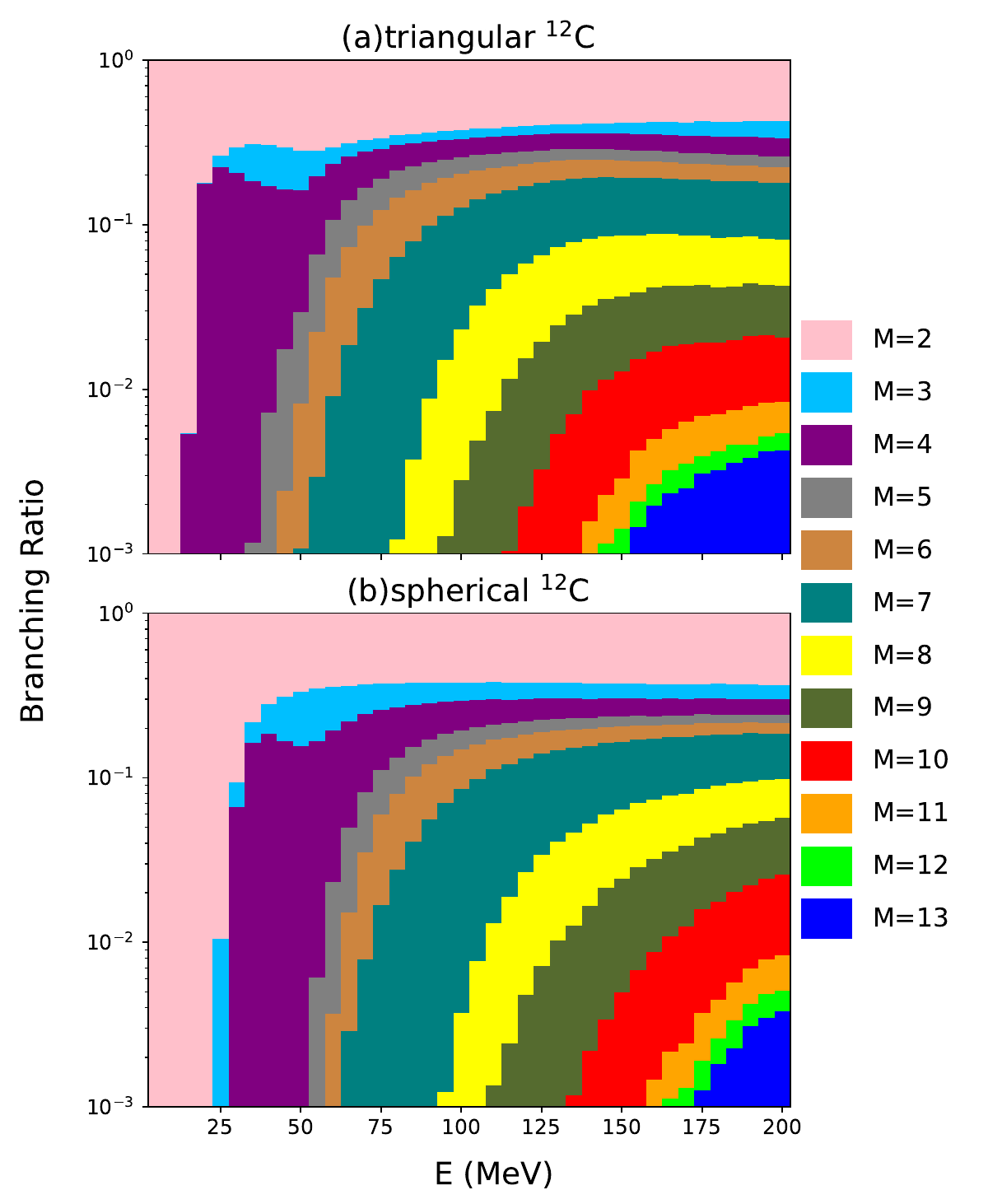}
  \caption{The distribution of the multiplicity branching ratio $p_i'$ of multiplicity from $M=2$ to $M=13$ as a function of the incident energy for the p + $^{12}$C reaction. Panel (a) shows the result of target $^{12}$C with triangular 3$\alpha$-cluster structure, while panel (b) shows the result of target $^{12}$C with spherical structure. Both panels are the results by the EQMD model with GEMINI decay process.}
  \label{fig_Msum}
\end{figure}

\subsection{Multiplicity branching ratio}\label{sec_br2}

Another kind of branching ratio, the multiplicity branching ratio $p_i'$, is also calculated, which is defined as,
\begin{equation}
  p_i' = \frac{N_i'}{N_0'}
  \label{eq_br}
\end{equation}
where $N_i'$ is the number of events with multiplicity $M=i$, and $N_0'$ is the number of all events. The fragment branching ratio distinguishes the types of emitted particles and fragments, i.e., the proton emission events and neutron emission events are treated as different types of events in the fragment branching ratio, while the multiplicity branching ratio does not distinguish the types of emitted particles or fragments and depends only on the number of emitted particles and fragments in each independent event. The multiplicity branching ratio is a more general observable than the fragment branching ratio and is more commonly used in heavy-ion collisions, which can reveal some macroscopic properties of the fragmentation systems, such as the liquid-gas phase transitions \cite{YGM1,Vinogradov2008BLPI}. Fig.~\ref{fig_Msum} shows the the multiplicity distribution for $M=2$ to $M=13$ as a function of the incident energy. The events with $M=2$ are mainly quasi-elastic reaction and have the largest branching ratio at all energies. The events with $M=3$ are mainly single-nucleon removal reactions, and the events with $M=4$ are the mixture of the 3-$\alpha$ decay of excited $^{12}$C and the two-nucleon removal reactions. The events with higher multiplicity $M>4$ reasonably distribute layer by layer like ripples.

\begin{figure}
  \includegraphics[width=\columnwidth]{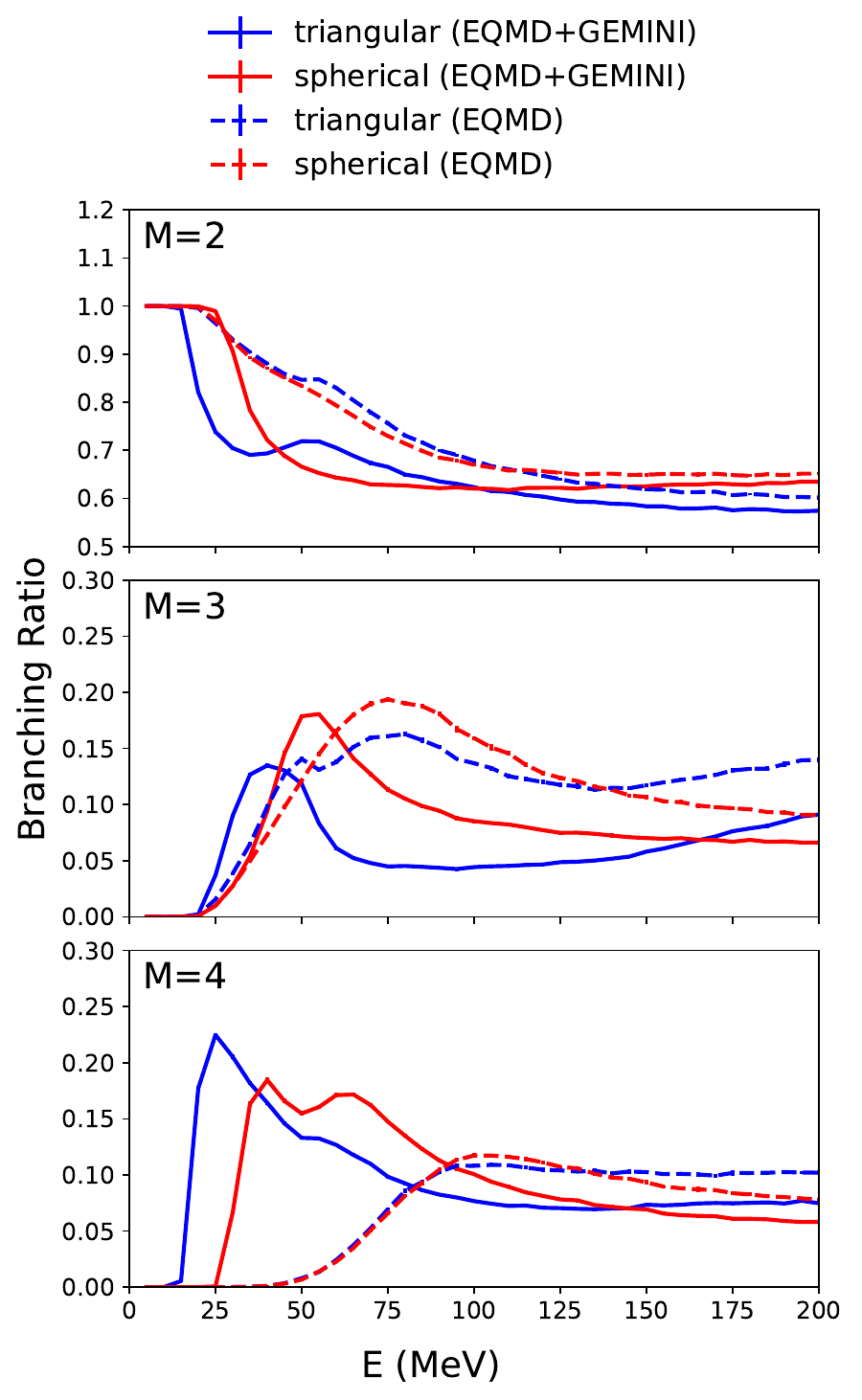}
  \caption{The incident energy dependence of the multiplicity branching ratio $p_i'$ of multiplicity from $M=2$, 3 and 4. The comparisons between the results with the EQMD only and the EQMD + GEMINI are given for both initial $^{12}$C structures.}
  \label{fig_br_mul}
\end{figure}

Fig.~\ref{fig_br_mul} shows the incident energy dependence of the multiplicity branching ratios for $M$ = 2, 3 or 4, respectively. The panel of $M=2$ is very similar to the panel of $^{12}$C in Fig.~\ref{fig_branch}. The panel of $M=3$ is the mixture of the panels of $^{11}$B and $^{11}$C in Fig.~\ref{fig_branch}, and the common feature of the panels of $^{11}$B and $^{11}$C also appears here. The panel of $M=4$ is mainly the mixture of the panels of $^4$He, $^{10}$B, $^{10}$Be and $^{10}$C in Fig.~\ref{fig_branch}. The lower energy part (about $E<50$ MeV) only relates to the 3-$\alpha$ decay of excited $^{12}$C, and the higher energy part (about $E>50$ MeV) is mixed with two-nucleon removal reactions. The features of these reaction channels discussed above can also be observed here.

\subsection{Fragment information entropy}

The event information entropy can be considered as a measurable value to evaluate the event-by-event fluctuation of a reaction system, and its value represents the complexity of the event distribution. The event information entropy can be calculated by Eq.~\ref{eq_ie} with different branching ratios, such as the fragment branching ratio $p_i$ and the multiplicity branching ratio $p_i'$ in this work.

\begin{figure}
  \includegraphics[width=\columnwidth]{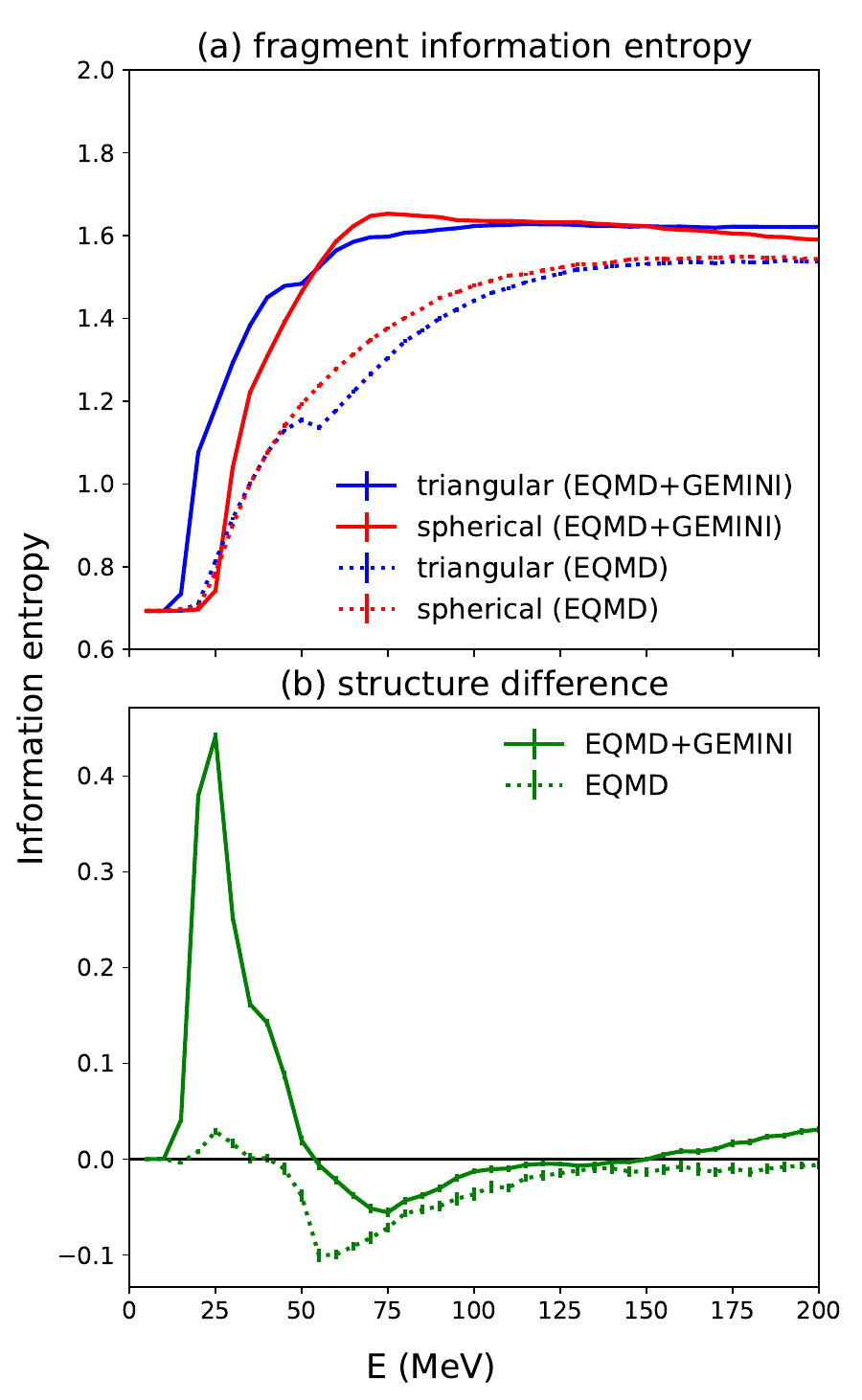}
  \caption{(a) The incident energy dependence of the fragment information entropy for the p+$^{12}$C reaction. The blue lines are the results of target $^{12}$C with triangular 3$\alpha$-cluster structure, while the red lines are the results of target $^{12}$C with spherical structure. The solid lines are the results by the EQMD model with GEMINI decay process, while the dashed lines are the results by only the EQMD model. (b) The difference in the fragment information entropy between the two structures, as a function of the incident energy. Specifically, $H_{\mathrm{diff}} = H_{\mathrm{tri}}-H_{\mathrm{sph}}$. For example, the green solid line is the result of the blue solid line minus the red solid line in panel (a).}
  \label{fig_ientropy}
\end{figure}

Fig.~\ref{fig_ientropy}(a) shows the incident energy dependence of the fragment information entropy, which is calculated by using the fragment branching ratios. Fig.~\ref{fig_ientropy}(b) shows the differences in fragment information entropy caused by different initial nuclear structures as a function of the incident energy. Let us first look at the dashed lines for direct dynamics results in panel (a). Compared to the result of spherical $^{12}$C (red dashed line), the result of triangular $^{12}$C (blue dashed line) shows a reduced component starting from $E\approx 50$ MeV and reaches a maximum reduction at $E\approx 55$ MeV. Such reduced component is displayed by the solid green line in Fig.~\ref{fig_ientropy}(b), and it is corresponding to the extra quasi-elastic cross section of the triangular $^{12}$C as discussed in Sec.~\ref{sec_br}. It is evident that the quasi-elastic reaction does not increase the chaoticity of the fragmentation system, and thus the fragment information entropy decreases as the branching ratio of quasi-elastic reaction increases. On the other hand, although less obvious, this feature can also be observed in the result after the decay process, i.e., there is a small dent at $E\approx 50$ MeV in the fragment information entropy of triangular $^{12}C$ (blue solid line) and a valley at $E\approx 75$ MeV in the structure difference (green solid line). The structure difference in the result after the decay process (green solid line) shows a high sharp peak at $E\approx 25$ MeV, which originates from the decay threshold shift due to the different binding energies of two $^{12}$C structures. We can imagine that if the blue solid line is shifted to the right side by about 15 MeV, the reduced component in the direct dynamics result (dashed lines) will also appear here in the solid lines.

\begin{figure}
  \includegraphics[width=\columnwidth]{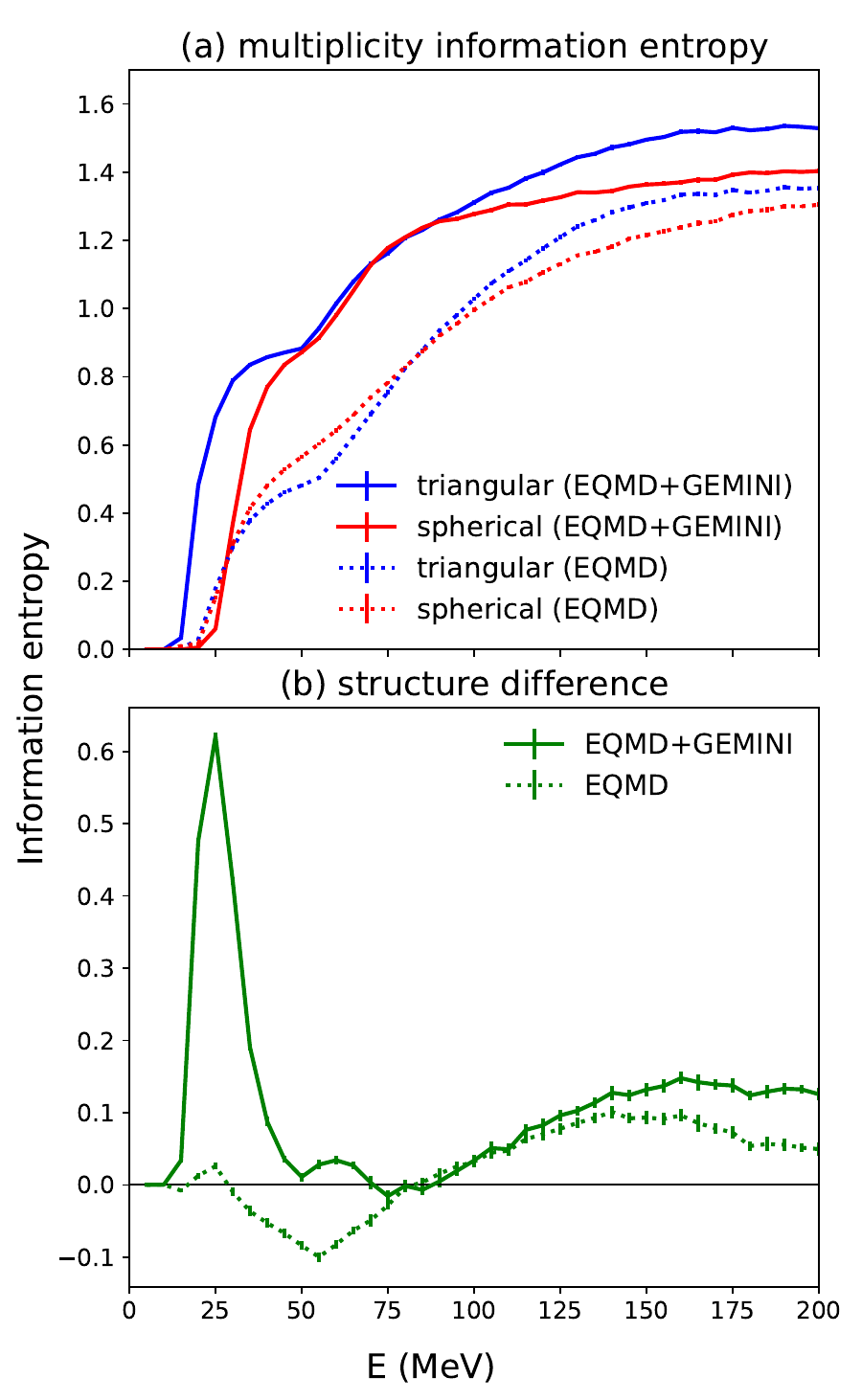}
  \caption{Same as Fig.~\ref{fig_ientropy} but for multiplicity information entropy.}
  \label{fig_mentropy}
\end{figure}

\subsection{ Multiplicity information entropy}
Fig.~\ref{fig_mentropy} shows the same calculation as Fig.~\ref{fig_ientropy} but by using the multiplicity branching ratios. As mentioned above, the multiplicity branching ratio has a more general definition than the fragment branching ratio, and thus the multiplicity information entropy is always much lower than the fragment information entropy. Fig.~\ref{fig_mentropy} shows a similar result to Fig.~\ref{fig_ientropy}, that the decayed result (blue solid line) and the direct dynamics result (blue dashed line) of triangular $^{12}$C both appear a dent at $E\approx50$ MeV compared to the results of spherical $^{12}$C (red lines). The result after the decay process in panel (b) (green dashed line) additionally shows a small peak at $E\approx 60$ MeV, which is probably due to the mixing of the 3-$\alpha$ decay and two-nucleon removal reactions.

\section{Conclusion}

In the present work we have used the EQMD model together with the GEMINI decay code to simulate the reaction of p + $^{12}$C with two different initial structures of $^{12}$C, namely triangular 3$\alpha$ structure and spherical structure, with the proton incident energies ranging from 5 to 200 MeV. Both results after the decay process and the direct dynamics result are discussed in this paper. The fragment branching ratios and multiplicity branching ratios are calculated, and they are discussed in details in Sec.~\ref{sec_br} and \ref{sec_br2}. The results indicate that compared to the spherical $^{12}$C, the triangular 3$\alpha$ $^{12}$C has an extra component of quasi-elastic reaction which is peaked at $E\approx 50$ MeV. The associated event information entropies, namely the fragment information entropy as well as the multiplicity information entropy, are calculated for these two structures. Both information entropies generally show an increasing trend with the incident energy. A small dent is observed in the information entropy curve for the triangular 3$\alpha$ $^{12}$C, which is corresponding to the extra component of quasi-elastic reaction for triangular 3$\alpha$ $^{12}$C. This special feature only appears in the curves of triangular 3$\alpha$ $^{12}$C according to the EQMD model with/without the decay process. Therefore, we propose that the event information entropy could be a probe for detecting the $\alpha$-cluster structure.

\section*{Acknowledgements}
This work is partially supported by the National Natural Science Foundation of China under Contracts No. $12147101$ and $11890714$, the Strategic Priority Research Program of CAS under Grant No.\ XDB34000000, the National Key R\&D Program of China under Grant No.\ 2022YFA1602300, the Guangdong Major Project of Basic and Applied Basic Research No.\ 2020B0301030008, the STCSM under Grant No.\ 23590780100, and the Natural Science Foundation of Shanghai under Grant No.\ 23JC1400200.

\end{CJK}
\bibliography{EQMD}

\begin{thebibliography}{43}%
\makeatletter
\providecommand \@ifxundefined [1]{%
 \@ifx{#1\undefined}
}%
\providecommand \@ifnum [1]{%
 \ifnum #1\expandafter \@firstoftwo
 \else \expandafter \@secondoftwo
 \fi
}%
\providecommand \@ifx [1]{%
 \ifx #1\expandafter \@firstoftwo
 \else \expandafter \@secondoftwo
 \fi
}%
\providecommand \natexlab [1]{#1}%
\providecommand \enquote  [1]{``#1''}%
\providecommand \bibnamefont  [1]{#1}%
\providecommand \bibfnamefont [1]{#1}%
\providecommand \citenamefont [1]{#1}%
\providecommand \href@noop [0]{\@secondoftwo}%
\providecommand \href [0]{\begingroup \@sanitize@url \@href}%
\providecommand \@href[1]{\@@startlink{#1}\@@href}%
\providecommand \@@href[1]{\endgroup#1\@@endlink}%
\providecommand \@sanitize@url [0]{\catcode `\\12\catcode `\$12\catcode
  `\&12\catcode `\#12\catcode `\^12\catcode `\_12\catcode `\%12\relax}%
\providecommand \@@startlink[1]{}%
\providecommand \@@endlink[0]{}%
\providecommand \url  [0]{\begingroup\@sanitize@url \@url }%
\providecommand \@url [1]{\endgroup\@href {#1}{\urlprefix }}%
\providecommand \urlprefix  [0]{URL }%
\providecommand \Eprint [0]{\href }%
\providecommand \doibase [0]{https://doi.org/}%
\providecommand \selectlanguage [0]{\@gobble}%
\providecommand \bibinfo  [0]{\@secondoftwo}%
\providecommand \bibfield  [0]{\@secondoftwo}%
\providecommand \translation [1]{[#1]}%
\providecommand \BibitemOpen [0]{}%
\providecommand \bibitemStop [0]{}%
\providecommand \bibitemNoStop [0]{.\EOS\space}%
\providecommand \EOS [0]{\spacefactor3000\relax}%
\providecommand \BibitemShut  [1]{\csname bibitem#1\endcsname}%
\let\auto@bib@innerbib\@empty
\bibitem [{\citenamefont {Cao}\ and\ \citenamefont {Hwa}(1995)}]{ZC1}%
  \BibitemOpen
  \bibfield  {author} {\bibinfo {author} {\bibfnamefont {Z.}~\bibnamefont
  {Cao}}\ and\ \bibinfo {author} {\bibfnamefont {R.~C.}\ \bibnamefont {Hwa}},\
  }\bibfield  {title} {\bibinfo {title} {In search for signs of chaos in
  branching processes},\ }\href {https://doi.org/10.1103/PhysRevLett.75.1268}
  {\bibfield  {journal} {\bibinfo  {journal} {Physical Review Letters}\
  }\textbf {\bibinfo {volume} {75}},\ \bibinfo {pages} {1268} (\bibinfo {year}
  {1995})}\BibitemShut {NoStop}%
\bibitem [{\citenamefont {Cao}\ and\ \citenamefont {Hwa}(1996)}]{ZC2}%
  \BibitemOpen
  \bibfield  {author} {\bibinfo {author} {\bibfnamefont {Z.}~\bibnamefont
  {Cao}}\ and\ \bibinfo {author} {\bibfnamefont {R.~C.}\ \bibnamefont {Hwa}},\
  }\bibfield  {title} {\bibinfo {title} {Chaotic behavior of particle
  production in branching processes},\ }\href
  {https://10.1103/physrevd.53.6608} {\bibfield  {journal} {\bibinfo  {journal}
  {Physical Review D}\ }\textbf {\bibinfo {volume} {53}},\ \bibinfo {pages}
  {6608} (\bibinfo {year} {1996})}\BibitemShut {NoStop}%
\bibitem [{\citenamefont {Brogueira}\ \emph {et~al.}(1996)\citenamefont
  {Brogueira}, \citenamefont {de~Deus},\ and\ \citenamefont {Da~Silva}}]{PB1}%
  \BibitemOpen
  \bibfield  {author} {\bibinfo {author} {\bibfnamefont {P.}~\bibnamefont
  {Brogueira}}, \bibinfo {author} {\bibfnamefont {J.~D.}\ \bibnamefont
  {de~Deus}},\ and\ \bibinfo {author} {\bibfnamefont {I.}~\bibnamefont
  {Da~Silva}},\ }\bibfield  {title} {\bibinfo {title} {Information entropy and
  particle production in branching processes},\ }\href
  {https://doi.org/10.1103/PhysRevD.53.5283} {\bibfield  {journal} {\bibinfo
  {journal} {Physical Review D}\ }\textbf {\bibinfo {volume} {53}},\ \bibinfo
  {pages} {5283} (\bibinfo {year} {1996})}\BibitemShut {NoStop}%
\bibitem [{\citenamefont {Ma}(1999)}]{YGM1}%
  \BibitemOpen
  \bibfield  {author} {\bibinfo {author} {\bibfnamefont {Y.-G.}\ \bibnamefont
  {Ma}},\ }\bibfield  {title} {\bibinfo {title} {Application of information
  theory in nuclear liquid gas phase transition},\ }\href
  {https://doi.org/10.1103/PhysRevLett.83.3617} {\bibfield  {journal} {\bibinfo
   {journal} {Physical Review Letters}\ }\textbf {\bibinfo {volume} {83}},\
  \bibinfo {pages} {3617} (\bibinfo {year} {1999})}\BibitemShut {NoStop}%
\bibitem [{\citenamefont {Ma}\ and\ \citenamefont {Ma}(2018)}]{CWM1}%
  \BibitemOpen
  \bibfield  {author} {\bibinfo {author} {\bibfnamefont {C.-W.}\ \bibnamefont
  {Ma}}\ and\ \bibinfo {author} {\bibfnamefont {Y.-G.}\ \bibnamefont {Ma}},\
  }\bibfield  {title} {\bibinfo {title} {Shannon information entropy in
  heavy-ion collisions},\ }\href {https://doi.org/10.1016/j.ppnp.2018.01.002}
  {\bibfield  {journal} {\bibinfo  {journal} {Progress in Particle and Nuclear
  Physics}\ }\textbf {\bibinfo {volume} {99}},\ \bibinfo {pages} {120}
  (\bibinfo {year} {2018})}\BibitemShut {NoStop}%
\bibitem [{\citenamefont {Shannon}(1948)}]{Shannon}%
  \BibitemOpen
  \bibfield  {author} {\bibinfo {author} {\bibfnamefont {C.~E.}\ \bibnamefont
  {Shannon}},\ }\bibfield  {title} {\bibinfo {title} {A mathematical theory of
  communication},\ }\href {https://doi.org/10.1002/j.1538-7305.1948.tb01338.x}
  {\bibfield  {journal} {\bibinfo  {journal} {The Bell System Technical
  Journal}\ }\textbf {\bibinfo {volume} {27}},\ \bibinfo {pages} {379}
  (\bibinfo {year} {1948})}\BibitemShut {NoStop}%
\bibitem [{\citenamefont {Ma}\ \emph {et~al.}(2003)\citenamefont {Ma},
  \citenamefont {Ma}, \citenamefont {Wang} \emph {et~al.}}]{GLM1}%
  \BibitemOpen
  \bibfield  {author} {\bibinfo {author} {\bibfnamefont {G.-L.}\ \bibnamefont
  {Ma}}, \bibinfo {author} {\bibfnamefont {Y.-G.}\ \bibnamefont {Ma}}, \bibinfo
  {author} {\bibfnamefont {K.}~\bibnamefont {Wang}}, \emph {et~al.},\
  }\bibfield  {title} {\bibinfo {title} {$\delta$-scaling and information
  entropy in ultra-relativistic nucleus-nucleus collisions},\ }\href
  {https://doi.org/10.1088/0256-307X/20/7/312} {\bibfield  {journal} {\bibinfo
  {journal} {Chinese Physics Letters}\ }\textbf {\bibinfo {volume} {20}},\
  \bibinfo {pages} {1013} (\bibinfo {year} {2003})}\BibitemShut {NoStop}%
\bibitem [{\citenamefont {Pu}\ \emph {et~al.}(2023)\citenamefont {Pu},
  \citenamefont {Yu}, \citenamefont {Cheng}, \citenamefont {Wang},
  \citenamefont {Guo},\ and\ \citenamefont {Ma}}]{JP1}%
  \BibitemOpen
  \bibfield  {author} {\bibinfo {author} {\bibfnamefont {J.}~\bibnamefont
  {Pu}}, \bibinfo {author} {\bibfnamefont {Y.-B.}\ \bibnamefont {Yu}}, \bibinfo
  {author} {\bibfnamefont {K.-X.}\ \bibnamefont {Cheng}}, \bibinfo {author}
  {\bibfnamefont {Y.-T.}\ \bibnamefont {Wang}}, \bibinfo {author}
  {\bibfnamefont {Y.-F.}\ \bibnamefont {Guo}},\ and\ \bibinfo {author}
  {\bibfnamefont {C.-W.}\ \bibnamefont {Ma}},\ }\bibfield  {title} {\bibinfo
  {title} {Exploring critical fluctuation phenomenon according to net-proton
  multiplicity information entropy in ampt model},\ }\href
  {https://doi.org/10.1016/j.physletb.2023.137909} {\bibfield  {journal}
  {\bibinfo  {journal} {Physics Letters B}\ }\textbf {\bibinfo {volume}
  {841}},\ \bibinfo {pages} {137909} (\bibinfo {year} {2023})}\BibitemShut
  {NoStop}%
\bibitem [{\citenamefont {Kou}\ and\ \citenamefont {Chen}(2023)}]{WK1}%
  \BibitemOpen
  \bibfield  {author} {\bibinfo {author} {\bibfnamefont {W.}~\bibnamefont
  {Kou}}\ and\ \bibinfo {author} {\bibfnamefont {X.}~\bibnamefont {Chen}},\
  }\bibfield  {title} {\bibinfo {title} {Mechanical structures inside proton
  with configurational entropy language},\ }\href
  {https://doi.org/10.1016/j.physletb.2023.138199} {\bibfield  {journal}
  {\bibinfo  {journal} {Physics Letters B}\ }\textbf {\bibinfo {volume}
  {846}},\ \bibinfo {pages} {138199} (\bibinfo {year} {2023})}\BibitemShut
  {NoStop}%
\bibitem [{\citenamefont {Deng}\ and\ \citenamefont {Ma}(2024)}]{Deng3}%
  \BibitemOpen
  \bibfield  {author} {\bibinfo {author} {\bibfnamefont {X.~G.}\ \bibnamefont
  {Deng}}\ and\ \bibinfo {author} {\bibfnamefont {Y.~G.}\ \bibnamefont {Ma}},\
  }\bibfield  {title} {\bibinfo {title} {Information entropy for central
  197au+197au collisions in the urqmd model.},\ }\href
  {https://doi.org/10.48550/arXiv.2404.03424} {\bibfield  {journal} {\bibinfo
  {journal} {arXiv}\ ,\ \bibinfo {pages} {2404.03424}} (\bibinfo {year}
  {2024})}\BibitemShut {NoStop}%
\bibitem [{\citenamefont {Ma}\ and\ \citenamefont {Ma}(2024)}]{MaWH}%
  \BibitemOpen
  \bibfield  {author} {\bibinfo {author} {\bibfnamefont {W.~H.}\ \bibnamefont
  {Ma}}\ and\ \bibinfo {author} {\bibfnamefont {Y.~G.}\ \bibnamefont {Ma}},\
  }\bibfield  {title} {\bibinfo {title} {Exploring the diversity of nuclear
  density through information entropy},\ }\href
  {https://doi.org/10.3390/e26090763} {\bibfield  {journal} {\bibinfo
  {journal} {Entropy}\ }\textbf {\bibinfo {volume} {26}},\ \bibinfo {pages}
  {763} (\bibinfo {year} {2024})}\BibitemShut {NoStop}%
\bibitem [{\citenamefont {Wolter}\ \emph {et~al.}(2022)\citenamefont {Wolter},
  \citenamefont {Colonna}, \citenamefont {Cozma}, \citenamefont {Danielewicz},
  \citenamefont {Ko}, \citenamefont {Kumar}, \citenamefont {Ono}, \citenamefont
  {Tsang}, \citenamefont {Xu}, \citenamefont {Zhang} \emph {et~al.}}]{Wolter}%
  \BibitemOpen
  \bibfield  {author} {\bibinfo {author} {\bibfnamefont {H.}~\bibnamefont
  {Wolter}}, \bibinfo {author} {\bibfnamefont {M.}~\bibnamefont {Colonna}},
  \bibinfo {author} {\bibfnamefont {D.}~\bibnamefont {Cozma}}, \bibinfo
  {author} {\bibfnamefont {P.}~\bibnamefont {Danielewicz}}, \bibinfo {author}
  {\bibfnamefont {C.~M.}\ \bibnamefont {Ko}}, \bibinfo {author} {\bibfnamefont
  {R.}~\bibnamefont {Kumar}}, \bibinfo {author} {\bibfnamefont
  {A.}~\bibnamefont {Ono}}, \bibinfo {author} {\bibfnamefont {M.~B.}\
  \bibnamefont {Tsang}}, \bibinfo {author} {\bibfnamefont {J.}~\bibnamefont
  {Xu}}, \bibinfo {author} {\bibfnamefont {Y.-X.}\ \bibnamefont {Zhang}}, \emph
  {et~al.},\ }\bibfield  {title} {\bibinfo {title} {Transport model comparison
  studies of intermediate-energy heavy-ion collisions},\ }\href
  {https://doi.org/10.1016/j.ppnp.2022.103962} {\bibfield  {journal} {\bibinfo
  {journal} {Progress in Particle and Nuclear Physics}\ }\textbf {\bibinfo
  {volume} {125}},\ \bibinfo {pages} {103962} (\bibinfo {year}
  {2022})}\BibitemShut {NoStop}%
\bibitem [{\citenamefont {Ono}\ and\ \citenamefont
  {Randrup}(2006)}]{Ono2006EPJA}%
  \BibitemOpen
  \bibfield  {author} {\bibinfo {author} {\bibfnamefont {A.}~\bibnamefont
  {Ono}}\ and\ \bibinfo {author} {\bibfnamefont {J.}~\bibnamefont {Randrup}},\
  }\bibfield  {title} {\bibinfo {title} {Dynamical models for fragment
  formation},\ }\bibfield  {booktitle} {\emph {\bibinfo {booktitle} {Dynamics
  and Thermodynamics with Nuclear Degrees of Freedom}},\ }\href
  {https://doi.org/10.1140/epja/i2006-10110-1} {\bibfield  {journal} {\bibinfo
  {journal} {Eur. Phys. J. A}\ }\textbf {\bibinfo {volume} {30}},\ \bibinfo
  {pages} {109} (\bibinfo {year} {2006})}\BibitemShut {NoStop}%
\bibitem [{\citenamefont {Ono}(2019)}]{Ono2019PPNP}%
  \BibitemOpen
  \bibfield  {author} {\bibinfo {author} {\bibfnamefont {A.}~\bibnamefont
  {Ono}},\ }\bibfield  {title} {\bibinfo {title} {Dynamics of clusters and
  fragments in heavy-ion collisions},\ }\href
  {https://doi.org/10.1016/j.ppnp.2018.11.001} {\bibfield  {journal} {\bibinfo
  {journal} {Progress in Particle and Nuclear Physics}\ }\textbf {\bibinfo
  {volume} {105}},\ \bibinfo {pages} {139} (\bibinfo {year}
  {2019})}\BibitemShut {NoStop}%
\bibitem [{\citenamefont {Xu}(2019)}]{Xu2019PPNP}%
  \BibitemOpen
  \bibfield  {author} {\bibinfo {author} {\bibfnamefont {J.}~\bibnamefont
  {Xu}},\ }\bibfield  {title} {\bibinfo {title} {Transport approaches for the
  description of intermediate-energy heavy-ion collisions},\ }\href
  {https://doi.org/10.1016/j.ppnp.2019.02.009} {\bibfield  {journal} {\bibinfo
  {journal} {Progress in Particle and Nuclear Physics}\ }\textbf {\bibinfo
  {volume} {106}},\ \bibinfo {pages} {312} (\bibinfo {year}
  {2019})}\BibitemShut {NoStop}%
\bibitem [{\citenamefont {Deng}\ \emph {et~al.}(2024)\citenamefont {Deng},
  \citenamefont {Fang},\ and\ \citenamefont {Ma}}]{Deng1}%
  \BibitemOpen
  \bibfield  {author} {\bibinfo {author} {\bibfnamefont {X.-G.}\ \bibnamefont
  {Deng}}, \bibinfo {author} {\bibfnamefont {D.-Q.}\ \bibnamefont {Fang}},\
  and\ \bibinfo {author} {\bibfnamefont {Y.-G.}\ \bibnamefont {Ma}},\
  }\bibfield  {title} {\bibinfo {title} {Shear viscosity of nucleonic matter},\
  }\href {https://doi.org/10.1016/j.ppnp.2023.104095} {\bibfield  {journal}
  {\bibinfo  {journal} {Progress in Particle and Nuclear Physics}\ }\textbf
  {\bibinfo {volume} {136}},\ \bibinfo {pages} {104095} (\bibinfo {year}
  {2024})}\BibitemShut {NoStop}%
\bibitem [{\citenamefont {Aichelin}(1991)}]{Aichelin}%
  \BibitemOpen
  \bibfield  {author} {\bibinfo {author} {\bibfnamefont {J.}~\bibnamefont
  {Aichelin}},\ }\bibfield  {title} {\bibinfo {title} {Quantum molecular
  dynamics - a dynamical microscopic n-body approach to investigate fragment
  formation and the nuclear equation of state in heavy ion collisions},\ }\href
  {https://doi.org/doi:10.1016/0370-1573(91)90094-3} {\bibfield  {journal}
  {\bibinfo  {journal} {Physics Reports}\ }\textbf {\bibinfo {volume} {202}},\
  \bibinfo {pages} {233} (\bibinfo {year} {1991})}\BibitemShut {NoStop}%
\bibitem [{\citenamefont {Deng}\ \emph {et~al.}(2022)\citenamefont {Deng},
  \citenamefont {Huang},\ and\ \citenamefont {Ma}}]{Deng2}%
  \BibitemOpen
  \bibfield  {author} {\bibinfo {author} {\bibfnamefont {X.-G.}\ \bibnamefont
  {Deng}}, \bibinfo {author} {\bibfnamefont {X.-G.}\ \bibnamefont {Huang}},\
  and\ \bibinfo {author} {\bibfnamefont {Y.-G.}\ \bibnamefont {Ma}},\
  }\bibfield  {title} {\bibinfo {title} {Lambda polarization in 108ag+ 108ag
  and 197au+ 197au collisions around a few gev},\ }\href
  {https://doi.org/10.1016/j.physletb.2022.137560} {\bibfield  {journal}
  {\bibinfo  {journal} {Physics Letters B}\ }\textbf {\bibinfo {volume}
  {835}},\ \bibinfo {pages} {137560} (\bibinfo {year} {2022})}\BibitemShut
  {NoStop}%
\bibitem [{\citenamefont {Kuttan}\ \emph {et~al.}(2023)\citenamefont {Kuttan},
  \citenamefont {Steinheimer}, \citenamefont {Zhou},\ and\ \citenamefont
  {Stoecker}}]{Zhou}%
  \BibitemOpen
  \bibfield  {author} {\bibinfo {author} {\bibfnamefont {M.~O.}\ \bibnamefont
  {Kuttan}}, \bibinfo {author} {\bibfnamefont {J.}~\bibnamefont {Steinheimer}},
  \bibinfo {author} {\bibfnamefont {K.}~\bibnamefont {Zhou}},\ and\ \bibinfo
  {author} {\bibfnamefont {H.}~\bibnamefont {Stoecker}},\ }\bibfield  {title}
  {\bibinfo {title} {Qcd equation of state of dense nuclear matter from a
  bayesian analysis of heavy-ion collision data},\ }\href
  {https://doi.org/10.1103/PhysRevLett.131.202303} {\bibfield  {journal}
  {\bibinfo  {journal} {Physical Review Letters}\ }\textbf {\bibinfo {volume}
  {131}},\ \bibinfo {pages} {202303} (\bibinfo {year} {2023})}\BibitemShut
  {NoStop}%
\bibitem [{\citenamefont {Qiao}\ \emph {et~al.}(2024)\citenamefont {Qiao},
  \citenamefont {Deng},\ and\ \citenamefont {Ma}}]{Qiao}%
  \BibitemOpen
  \bibfield  {author} {\bibinfo {author} {\bibfnamefont {F.-H.}\ \bibnamefont
  {Qiao}}, \bibinfo {author} {\bibfnamefont {X.-G.}\ \bibnamefont {Deng}},\
  and\ \bibinfo {author} {\bibfnamefont {Y.-G.}\ \bibnamefont {Ma}},\
  }\bibfield  {title} {\bibinfo {title} {Momentum correlation of light nuclei
  in au+ au collisions at gev},\ }\href
  {https://doi.org/10.1016/j.physletb.2024.138535} {\bibfield  {journal}
  {\bibinfo  {journal} {Physics Letters B}\ }\textbf {\bibinfo {volume}
  {850}},\ \bibinfo {pages} {138535} (\bibinfo {year} {2024})}\BibitemShut
  {NoStop}%
\bibitem [{\citenamefont {Ding}\ \emph
  {et~al.}(2024{\natexlab{a}})\citenamefont {Ding}, \citenamefont {Fang},\ and\
  \citenamefont {Ma}}]{Ding}%
  \BibitemOpen
  \bibfield  {author} {\bibinfo {author} {\bibfnamefont {M.-Q.}\ \bibnamefont
  {Ding}}, \bibinfo {author} {\bibfnamefont {D.-Q.}\ \bibnamefont {Fang}},\
  and\ \bibinfo {author} {\bibfnamefont {Y.-G.}\ \bibnamefont {Ma}},\
  }\bibfield  {title} {\bibinfo {title} {Effects of neutron-skin thickness on
  light-particle production},\ }\href
  {https://doi.org/10.1103/PhysRevC.109.024616} {\bibfield  {journal} {\bibinfo
   {journal} {Physical Review C}\ }\textbf {\bibinfo {volume} {109}},\ \bibinfo
  {pages} {024616} (\bibinfo {year} {2024}{\natexlab{a}})}\BibitemShut
  {NoStop}%
\bibitem [{\citenamefont {Buyukcizmeci}\ \emph {et~al.}(2023)\citenamefont
  {Buyukcizmeci}, \citenamefont {Reichert}, \citenamefont {Botvina},\ and\
  \citenamefont {Bleicher}}]{Buy}%
  \BibitemOpen
  \bibfield  {author} {\bibinfo {author} {\bibfnamefont {N.}~\bibnamefont
  {Buyukcizmeci}}, \bibinfo {author} {\bibfnamefont {T.}~\bibnamefont
  {Reichert}}, \bibinfo {author} {\bibfnamefont {A.}~\bibnamefont {Botvina}},\
  and\ \bibinfo {author} {\bibfnamefont {M.}~\bibnamefont {Bleicher}},\
  }\bibfield  {title} {\bibinfo {title} {Nucleosynthesis of light nuclei and
  hypernuclei in central au+ au collisions at $\sqrt{s_{NN}}$= 3 gev},\ }\href
  {https://doi.org/10.1103/PhysRevC.108.054904} {\bibfield  {journal} {\bibinfo
   {journal} {Physical Review C}\ }\textbf {\bibinfo {volume} {108}},\ \bibinfo
  {pages} {054904} (\bibinfo {year} {2023})}\BibitemShut {NoStop}%
\bibitem [{\citenamefont {Cao}\ \emph {et~al.}(2023)\citenamefont {Cao},
  \citenamefont {Deng},\ and\ \citenamefont {Ma}}]{Cao}%
  \BibitemOpen
  \bibfield  {author} {\bibinfo {author} {\bibfnamefont {Y.~T.}\ \bibnamefont
  {Cao}}, \bibinfo {author} {\bibfnamefont {X.~G.}\ \bibnamefont {Deng}},\ and\
  \bibinfo {author} {\bibfnamefont {Y.~G.}\ \bibnamefont {Ma}},\ }\bibfield
  {title} {\bibinfo {title} {Impact of magnetic field on the giant dipole
  resonance of 40 ca using an extended quantum molecular dynamics model},\
  }\href {https://doi.org/10.1103/PhysRevC.108.049904} {\bibfield  {journal}
  {\bibinfo  {journal} {Physical Review C}\ }\textbf {\bibinfo {volume}
  {108}},\ \bibinfo {pages} {049904} (\bibinfo {year} {2023})}\BibitemShut
  {NoStop}%
\bibitem [{\citenamefont {Wang}\ \emph
  {et~al.}(2023{\natexlab{a}})\citenamefont {Wang}, \citenamefont {Ma},
  \citenamefont {He}, \citenamefont {Fang}, \citenamefont {Cao} \emph
  {et~al.}}]{WangSS}%
  \BibitemOpen
  \bibfield  {author} {\bibinfo {author} {\bibfnamefont {S.~S.}\ \bibnamefont
  {Wang}}, \bibinfo {author} {\bibfnamefont {Y.~G.}\ \bibnamefont {Ma}},
  \bibinfo {author} {\bibfnamefont {W.~B.}\ \bibnamefont {He}}, \bibinfo
  {author} {\bibfnamefont {D.~Q.}\ \bibnamefont {Fang}}, \bibinfo {author}
  {\bibfnamefont {X.~G.}\ \bibnamefont {Cao}}, \emph {et~al.},\ }\bibfield
  {title} {\bibinfo {title} {Influences of $\alpha$-clustering configurations
  on the giant dipole resonance in hot compound systems},\ }\href
  {https://doi.org/10.1103/PhysRevC.108.014609} {\bibfield  {journal} {\bibinfo
   {journal} {Physical Review C}\ }\textbf {\bibinfo {volume} {108}},\ \bibinfo
  {pages} {014609} (\bibinfo {year} {2023}{\natexlab{a}})}\BibitemShut
  {NoStop}%
\bibitem [{\citenamefont {Nara}\ \emph {et~al.}(2022)\citenamefont {Nara},
  \citenamefont {Jinno}, \citenamefont {Murase},\ and\ \citenamefont
  {Ohnishi}}]{Nara}%
  \BibitemOpen
  \bibfield  {author} {\bibinfo {author} {\bibfnamefont {Y.}~\bibnamefont
  {Nara}}, \bibinfo {author} {\bibfnamefont {A.}~\bibnamefont {Jinno}},
  \bibinfo {author} {\bibfnamefont {K.}~\bibnamefont {Murase}},\ and\ \bibinfo
  {author} {\bibfnamefont {A.}~\bibnamefont {Ohnishi}},\ }\bibfield  {title}
  {\bibinfo {title} {Directed flow of $\lambda$ in high-energy heavy-ion
  collisions and $\lambda$ potential in dense nuclear matter},\ }\href
  {https://doi.org/10.1103/PhysRevC.106.044902} {\bibfield  {journal} {\bibinfo
   {journal} {Physical Review C}\ }\textbf {\bibinfo {volume} {106}},\ \bibinfo
  {pages} {044902} (\bibinfo {year} {2022})}\BibitemShut {NoStop}%
\bibitem [{\citenamefont {Wang}\ \emph
  {et~al.}(2022{\natexlab{a}})\citenamefont {Wang}, \citenamefont {Ma},
  \citenamefont {Fang}, \citenamefont {Liu} \emph {et~al.}}]{WangTT}%
  \BibitemOpen
  \bibfield  {author} {\bibinfo {author} {\bibfnamefont {T.-T.}\ \bibnamefont
  {Wang}}, \bibinfo {author} {\bibfnamefont {Y.-G.}\ \bibnamefont {Ma}},
  \bibinfo {author} {\bibfnamefont {D.-Q.}\ \bibnamefont {Fang}}, \bibinfo
  {author} {\bibfnamefont {H.-L.}\ \bibnamefont {Liu}}, \emph {et~al.},\
  }\bibfield  {title} {\bibinfo {title} {Temperature and density effects on the
  two-nucleon momentum correlation function from excited single nuclei},\
  }\href {https://doi.org/10.1103/PhysRevC.105.024620} {\bibfield  {journal}
  {\bibinfo  {journal} {Physical Review C}\ }\textbf {\bibinfo {volume}
  {105}},\ \bibinfo {pages} {024620} (\bibinfo {year}
  {2022}{\natexlab{a}})}\BibitemShut {NoStop}%
\bibitem [{\citenamefont {Wang}\ \emph
  {et~al.}(2023{\natexlab{b}})\citenamefont {Wang}, \citenamefont {Yang},
  \citenamefont {Chen}, \citenamefont {Cui}, \citenamefont {Wang},
  \citenamefont {Xiao}, \citenamefont {Li},\ and\ \citenamefont
  {Zhang}}]{WangFY}%
  \BibitemOpen
  \bibfield  {author} {\bibinfo {author} {\bibfnamefont {F.-Y.}\ \bibnamefont
  {Wang}}, \bibinfo {author} {\bibfnamefont {J.-P.}\ \bibnamefont {Yang}},
  \bibinfo {author} {\bibfnamefont {X.}~\bibnamefont {Chen}}, \bibinfo {author}
  {\bibfnamefont {Y.}~\bibnamefont {Cui}}, \bibinfo {author} {\bibfnamefont
  {Y.-J.}\ \bibnamefont {Wang}}, \bibinfo {author} {\bibfnamefont {Z.-G.}\
  \bibnamefont {Xiao}}, \bibinfo {author} {\bibfnamefont {Z.-X.}\ \bibnamefont
  {Li}},\ and\ \bibinfo {author} {\bibfnamefont {Y.-X.}\ \bibnamefont
  {Zhang}},\ }\bibfield  {title} {\bibinfo {title} {Probing nucleon effective
  mass splitting with light particle emission},\ }\href
  {https://doi.org/10.1007/s41365-023-01241-z} {\bibfield  {journal} {\bibinfo
  {journal} {Nuclear Science and Techniques}\ }\textbf {\bibinfo {volume}
  {34}},\ \bibinfo {pages} {94} (\bibinfo {year}
  {2023}{\natexlab{b}})}\BibitemShut {NoStop}%
\bibitem [{\citenamefont {Xiao}\ \emph {et~al.}(2023)\citenamefont {Xiao},
  \citenamefont {Li}, \citenamefont {Wang}, \citenamefont {Liu},\ and\
  \citenamefont {Li}}]{XiaoK}%
  \BibitemOpen
  \bibfield  {author} {\bibinfo {author} {\bibfnamefont {K.}~\bibnamefont
  {Xiao}}, \bibinfo {author} {\bibfnamefont {P.-C.}\ \bibnamefont {Li}},
  \bibinfo {author} {\bibfnamefont {Y.-J.}\ \bibnamefont {Wang}}, \bibinfo
  {author} {\bibfnamefont {F.-H.}\ \bibnamefont {Liu}},\ and\ \bibinfo {author}
  {\bibfnamefont {Q.-F.}\ \bibnamefont {Li}},\ }\bibfield  {title} {\bibinfo
  {title} {Effects of sequential decay on collective flows and nuclear stopping
  power in heavy-ion collisions at intermediate energies},\ }\href
  {https://doi.org/10.1007/s41365-023-01205-3} {\bibfield  {journal} {\bibinfo
  {journal} {Nuclear Science and Techniques}\ }\textbf {\bibinfo {volume}
  {34}},\ \bibinfo {pages} {62} (\bibinfo {year} {2023})}\BibitemShut {NoStop}%
\bibitem [{\citenamefont {Wang}\ \emph
  {et~al.}(2022{\natexlab{b}})\citenamefont {Wang}, \citenamefont {Ou},\ and\
  \citenamefont {Xiao}}]{WangRS}%
  \BibitemOpen
  \bibfield  {author} {\bibinfo {author} {\bibfnamefont {R.-S.}\ \bibnamefont
  {Wang}}, \bibinfo {author} {\bibfnamefont {L.}~\bibnamefont {Ou}},\ and\
  \bibinfo {author} {\bibfnamefont {Z.-G.}\ \bibnamefont {Xiao}},\ }\bibfield
  {title} {\bibinfo {title} {Production of high-energy neutron beam from
  deuteron breakup},\ }\href {https://doi.org/10.1007/s41365-022-01075-1}
  {\bibfield  {journal} {\bibinfo  {journal} {Nuclear Science and Techniques}\
  }\textbf {\bibinfo {volume} {33}},\ \bibinfo {pages} {92} (\bibinfo {year}
  {2022}{\natexlab{b}})}\BibitemShut {NoStop}%
\bibitem [{\citenamefont {Liu}\ \emph {et~al.}(2022)\citenamefont {Liu},
  \citenamefont {Deng},\ and\ \citenamefont {Ma}}]{LiuC}%
  \BibitemOpen
  \bibfield  {author} {\bibinfo {author} {\bibfnamefont {C.}~\bibnamefont
  {Liu}}, \bibinfo {author} {\bibfnamefont {X.-G.}\ \bibnamefont {Deng}},\ and\
  \bibinfo {author} {\bibfnamefont {Y.-G.}\ \bibnamefont {Ma}},\ }\bibfield
  {title} {\bibinfo {title} {Density fluctuations in intermediate-energy
  heavy-ion collisions},\ }\href {https://doi.org/10.1007/s41365-022-01040-y}
  {\bibfield  {journal} {\bibinfo  {journal} {Nuclear Science and Techniques}\
  }\textbf {\bibinfo {volume} {33}},\ \bibinfo {pages} {52} (\bibinfo {year}
  {2022})}\BibitemShut {NoStop}%
\bibitem [{\citenamefont {Li}\ \emph {et~al.}(2022)\citenamefont {Li},
  \citenamefont {Wang},\ and\ \citenamefont {Zhang}}]{LiL}%
  \BibitemOpen
  \bibfield  {author} {\bibinfo {author} {\bibfnamefont {L.}~\bibnamefont
  {Li}}, \bibinfo {author} {\bibfnamefont {F.-Y.}\ \bibnamefont {Wang}},\ and\
  \bibinfo {author} {\bibfnamefont {Y.-X.}\ \bibnamefont {Zhang}},\ }\bibfield
  {title} {\bibinfo {title} {Isospin effects on intermediate mass fragments at
  intermediate energy-heavy ion collisions},\ }\href
  {https://doi.org/10.1007/s41365-022-01050-w} {\bibfield  {journal} {\bibinfo
  {journal} {Nuclear Science and Techniques}\ }\textbf {\bibinfo {volume}
  {33}},\ \bibinfo {pages} {58} (\bibinfo {year} {2022})}\BibitemShut {NoStop}%
\bibitem [{\citenamefont {Li}\ \emph {et~al.}(2023{\natexlab{a}})\citenamefont
  {Li}, \citenamefont {Wang}, \citenamefont {Li},\ and\ \citenamefont
  {Zhang}}]{LiPC1}%
  \BibitemOpen
  \bibfield  {author} {\bibinfo {author} {\bibfnamefont {P.}~\bibnamefont
  {Li}}, \bibinfo {author} {\bibfnamefont {Y.}~\bibnamefont {Wang}}, \bibinfo
  {author} {\bibfnamefont {Q.}~\bibnamefont {Li}},\ and\ \bibinfo {author}
  {\bibfnamefont {H.}~\bibnamefont {Zhang}},\ }\bibfield  {title} {\bibinfo
  {title} {Transport model analysis of the pion interferometry in au+ au
  collisions at e beam= 1.23 gev/nucleon},\ }\href
  {https://doi.org/10.1007/s11433-022-2026-5} {\bibfield  {journal} {\bibinfo
  {journal} {Science China Physics, Mechanics and Astronomy}\ }\textbf
  {\bibinfo {volume} {66}},\ \bibinfo {pages} {222011} (\bibinfo {year}
  {2023}{\natexlab{a}})}\BibitemShut {NoStop}%
\bibitem [{\citenamefont {Li}\ \emph {et~al.}(2023{\natexlab{b}})\citenamefont
  {Li}, \citenamefont {Steinheimer}, \citenamefont {Reichert}, \citenamefont
  {Kittiratpattana}, \citenamefont {Bleicher},\ and\ \citenamefont
  {Li}}]{LiPC2}%
  \BibitemOpen
  \bibfield  {author} {\bibinfo {author} {\bibfnamefont {P.}~\bibnamefont
  {Li}}, \bibinfo {author} {\bibfnamefont {J.}~\bibnamefont {Steinheimer}},
  \bibinfo {author} {\bibfnamefont {T.}~\bibnamefont {Reichert}}, \bibinfo
  {author} {\bibfnamefont {A.}~\bibnamefont {Kittiratpattana}}, \bibinfo
  {author} {\bibfnamefont {M.}~\bibnamefont {Bleicher}},\ and\ \bibinfo
  {author} {\bibfnamefont {Q.}~\bibnamefont {Li}},\ }\bibfield  {title}
  {\bibinfo {title} {Effects of a phase transition on two-pion interferometry
  in heavy ion collisions at s nn= 2.4- 7.7 gev},\ }\href
  {https://doi.org/10.1007/s11433-022-2041-8} {\bibfield  {journal} {\bibinfo
  {journal} {Science China Physics, Mechanics and Astronomy}\ }\textbf
  {\bibinfo {volume} {66}},\ \bibinfo {pages} {232011} (\bibinfo {year}
  {2023}{\natexlab{b}})}\BibitemShut {NoStop}%
\bibitem [{\citenamefont {Ding}\ \emph
  {et~al.}(2024{\natexlab{b}})\citenamefont {Ding}, \citenamefont {Fang},\ and\
  \citenamefont {Ma}}]{Ding2}%
  \BibitemOpen
  \bibfield  {author} {\bibinfo {author} {\bibfnamefont {M.-Q.}\ \bibnamefont
  {Ding}}, \bibinfo {author} {\bibfnamefont {D.-Q.}\ \bibnamefont {Fang}},\
  and\ \bibinfo {author} {\bibfnamefont {Y.-G.}\ \bibnamefont {Ma}},\
  }\bibfield  {title} {\bibinfo {title} {Neutron skin and its effects in
  heavy-ion collisions},\ }\href {https://doi.org/10.48550/arXiv.2409.07059}
  {\bibfield  {journal} {\bibinfo  {journal} {arXiv}\ ,\ \bibinfo {pages}
  {2409.07059}} (\bibinfo {year} {2024}{\natexlab{b}})}\BibitemShut {NoStop}%
\bibitem [{\citenamefont {Maruyama}\ \emph {et~al.}(1996)\citenamefont
  {Maruyama}, \citenamefont {Niita},\ and\ \citenamefont {Iwamoto}}]{Maruyama}%
  \BibitemOpen
  \bibfield  {author} {\bibinfo {author} {\bibfnamefont {T.}~\bibnamefont
  {Maruyama}}, \bibinfo {author} {\bibfnamefont {K.}~\bibnamefont {Niita}},\
  and\ \bibinfo {author} {\bibfnamefont {A.}~\bibnamefont {Iwamoto}},\
  }\bibfield  {title} {\bibinfo {title} {Extension of quantum molecular
  dynamics and its application to heavy-ion collisions},\ }\href
  {https://doi.org/10.1103/PhysRevC.53.297} {\bibfield  {journal} {\bibinfo
  {journal} {Physical Review C}\ }\textbf {\bibinfo {volume} {53}},\ \bibinfo
  {pages} {297} (\bibinfo {year} {1996})}\BibitemShut {NoStop}%
\bibitem [{\citenamefont {He}\ \emph {et~al.}(2014)\citenamefont {He},
  \citenamefont {Ma}, \citenamefont {Cao}, \citenamefont {Cai},\ and\
  \citenamefont {Zhang}}]{W.B.He}%
  \BibitemOpen
  \bibfield  {author} {\bibinfo {author} {\bibfnamefont {W.~B.}\ \bibnamefont
  {He}}, \bibinfo {author} {\bibfnamefont {Y.~G.}\ \bibnamefont {Ma}}, \bibinfo
  {author} {\bibfnamefont {X.~G.}\ \bibnamefont {Cao}}, \bibinfo {author}
  {\bibfnamefont {X.~Z.}\ \bibnamefont {Cai}},\ and\ \bibinfo {author}
  {\bibfnamefont {G.~Q.}\ \bibnamefont {Zhang}},\ }\bibfield  {title} {\bibinfo
  {title} {Giant dipole resonance as a fingerprint of $\alpha$ clustering
  configurations in c 12 and o 16},\ }\href
  {https://doi.org/10.1103/PhysRevLett.113.032506} {\bibfield  {journal}
  {\bibinfo  {journal} {Physical Review Letters}\ }\textbf {\bibinfo {volume}
  {113}},\ \bibinfo {pages} {032506} (\bibinfo {year} {2014})}\BibitemShut
  {NoStop}%
\bibitem [{\citenamefont {He}\ \emph {et~al.}(2016)\citenamefont {He},
  \citenamefont {Ma}, \citenamefont {Cao}, \citenamefont {Cai},\ and\
  \citenamefont {Zhang}}]{W.B.He2}%
  \BibitemOpen
  \bibfield  {author} {\bibinfo {author} {\bibfnamefont {W.~B.}\ \bibnamefont
  {He}}, \bibinfo {author} {\bibfnamefont {Y.~G.}\ \bibnamefont {Ma}}, \bibinfo
  {author} {\bibfnamefont {X.~G.}\ \bibnamefont {Cao}}, \bibinfo {author}
  {\bibfnamefont {X.~Z.}\ \bibnamefont {Cai}},\ and\ \bibinfo {author}
  {\bibfnamefont {G.~Q.}\ \bibnamefont {Zhang}},\ }\bibfield  {title} {\bibinfo
  {title} {Dipole oscillation modes in light $\alpha$-clustering nuclei},\
  }\href {http://dx.doi.org/10.1103/PhysRevC.94.014301} {\bibfield  {journal}
  {\bibinfo  {journal} {Physical Review C}\ }\textbf {\bibinfo {volume} {94}},\
  \bibinfo {pages} {014301} (\bibinfo {year} {2016})}\BibitemShut {NoStop}%
\bibitem [{\citenamefont {Huang}\ and\ \citenamefont {Ma}(2021)}]{Huang2021}%
  \BibitemOpen
  \bibfield  {author} {\bibinfo {author} {\bibfnamefont {B.-S.}\ \bibnamefont
  {Huang}}\ and\ \bibinfo {author} {\bibfnamefont {Y.-G.}\ \bibnamefont {Ma}},\
  }\bibfield  {title} {\bibinfo {title} {Dipole excitation of li 6 and be 9
  studied with an extended quantum molecular dynamics model},\ }\href
  {https://doi.org/10.1103/PhysRevC.103.054318} {\bibfield  {journal} {\bibinfo
   {journal} {Physical Review C}\ }\textbf {\bibinfo {volume} {103}},\ \bibinfo
  {pages} {054318} (\bibinfo {year} {2021})}\BibitemShut {NoStop}%
\bibitem [{\citenamefont {Chen}\ \emph {et~al.}(2023)\citenamefont {Chen},
  \citenamefont {Ye}, \citenamefont {Wei} \emph {et~al.}}]{YeYL}%
  \BibitemOpen
  \bibfield  {author} {\bibinfo {author} {\bibfnamefont {Y.}~\bibnamefont
  {Chen}}, \bibinfo {author} {\bibfnamefont {Y.-L.}\ \bibnamefont {Ye}},
  \bibinfo {author} {\bibfnamefont {K.}~\bibnamefont {Wei}}, \emph {et~al.},\
  }\bibfield  {title} {\bibinfo {title} {Progress and perspective of the
  research on exotic structures of unstable nuclei},\ }\href
  {https://doi.org/10.11889/j.0253-3219.2023.hjs.46.080020} {\bibfield
  {journal} {\bibinfo  {journal} {Nuclear Techniques (in Chinese)}\ }\textbf
  {\bibinfo {volume} {46}},\ \bibinfo {pages} {080020} (\bibinfo {year}
  {2023})}\BibitemShut {NoStop}%
\bibitem [{\citenamefont {Ma}(2023)}]{MaYG}%
  \BibitemOpen
  \bibfield  {author} {\bibinfo {author} {\bibfnamefont {Y.-G.}\ \bibnamefont
  {Ma}},\ }\bibfield  {title} {\bibinfo {title} {Effects of $\alpha$-clustering
  structure on nuclear reaction and relativistic heavy-ion collisions},\ }\href
  {https://doi.org/10.11889/j.0253-3219.2023.hjs.46.080001} {\bibfield
  {journal} {\bibinfo  {journal} {Nuclear Techniques (in Chinese)}\ }\textbf
  {\bibinfo {volume} {46}},\ \bibinfo {pages} {080001} (\bibinfo {year}
  {2023})}\BibitemShut {NoStop}%
\bibitem [{\citenamefont {Ma}\ and\ \citenamefont {Zhang}(2024)}]{MaYG2}%
  \BibitemOpen
  \bibfield  {author} {\bibinfo {author} {\bibfnamefont {Y.-G.}\ \bibnamefont
  {Ma}}\ and\ \bibinfo {author} {\bibfnamefont {S.}~\bibnamefont {Zhang}},\
  }\bibfield  {title} {\bibinfo {title} {$\alpha$-clustering effects in
  relativistic heavy-ion collisions (in chinese)},\ }\href
  {https://doi.org/10.1360/SSPMA-2024-0013} {\bibfield  {journal} {\bibinfo
  {journal} {SCIENTIA SINICA Physica, Mechanica and Astronomica (in Chinese)}\
  }\textbf {\bibinfo {volume} {54}},\ \bibinfo {pages} {292004} (\bibinfo
  {year} {2024})}\BibitemShut {NoStop}%
\bibitem [{\citenamefont {Charity}(2008)}]{Charity2008INDC}%
  \BibitemOpen
  \bibfield  {author} {\bibinfo {author} {\bibfnamefont {R.~J.}\ \bibnamefont
  {Charity}},\ }\bibfield  {title} {\bibinfo {title} {Gemini: A code to
  simulate the decay of a compound nucleus by a series of binary decays},\
  }\href@noop {} {\bibfield  {journal} {\bibinfo  {journal} {International
  Nuclear Data Committee}\ }\textbf {\bibinfo {volume} {0530}},\ \bibinfo
  {pages} {139} (\bibinfo {year} {2008})}\BibitemShut {NoStop}%
\bibitem [{\citenamefont {Vinogradov}(2008)}]{Vinogradov2008BLPI}%
  \BibitemOpen
  \bibfield  {author} {\bibinfo {author} {\bibfnamefont {A.~V.}\ \bibnamefont
  {Vinogradov}},\ }\bibfield  {title} {\bibinfo {title} {Qcd nuclear factor and
  moments of multiplicity distributions in higher orders of perturbation
  theory},\ }\href {https://doi.org/10.3103/S1068335608050011} {\bibfield
  {journal} {\bibinfo  {journal} {Bulletin of the Lebedev Physics Institute}\
  }\textbf {\bibinfo {volume} {35}},\ \bibinfo {pages} {131} (\bibinfo {year}
  {2008})}\BibitemShut {NoStop}%
\end{thebibliography}%
\end{document}